\documentclass[print]{aa}  

\usepackage{graphicx}
\usepackage{txfonts}
\begin{document}

   \title{Episodic accretion in binary protostars emerging from self-gravitating solar mass cores}

   
   \titlerunning{Episodic accretion in VLMS}

   \author{R. Riaz
          \inst{1}
          \and
          S. Vanaverbeke\inst{2}
          \and
          D.R.G. Schleicher\inst{1}
          }

   \institute{Departamento de Astronom\'ia, Facultad Ciencias F\'isicas y Matem\'aticas, Universidad de Concepci\'on, Av. Esteban Iturra s/n Barrio \\
Universitario, Casilla $160$-C, Concepci\'on, Chile\\
              \email{rriaz@astro-udec.cl}
              \email{dschleicher@astro-udec.cl}
         \and
             Centre for mathematical Plasma-Astrophysics, Department of Mathematics, KU Leuven, Celestijnenlaan 200B, 3001 Heverlee, Belgium\\
             \email{siegfriedvanaverbeke@gmail.com}
             }

   \date{Received ***; accepted ***}

 
  \abstract
   {Observations show a large spread in the luminosities of young protostars, which are frequently explained in the context of episodic accretion. We here test this scenario using numerical simulations following the collapse of a solar mass molecular cloud using the GRADSPH code, varying the strength of the initial perturbations and the temperature of the cores. A specific emphasis of this paper is to investigate the role of binaries and multiple systems in the context of episodic accretion, and to compare their evolution to the evolution in isolated fragments. Our models form a variety of low mass protostellar objects including single, binary and triple systems with binaries more active in exhibiting episodic accretion than isolated protostars. We also find a general decreasing trend for the average mass accretion rate over time, suggesting that the majority of the protostellar mass is accreted within the first 10$^{5}$ years. This result can potentially help to explain the surprisingly low average luminosities in the majority of the protostellar population.}


   \maketitle
%

\section{Introduction}

  Understanding the non-linear dynamics of star formation poses a significant challenge. During protostellar collapse, the gas which is part of the interstellar medium experiences a more than 10 orders of magnitude increase in density. However, despite the challenges, non-linear phenomena such as episodic accretion can potentially explain some observed inconsistencies which are related to the luminosity of young low-mass protostars \citep{b1}. For example, if there is a constant flow of matter from a collapsing envelope of gas onto protostars we would expect that a sample of these low-mass objects should have the same luminosities, but instead observations have shown that low-mass protostars indeed do have different luminosities \citep{b70}. In fact, a range of observed protostellar luminosities distribution from 0.01 $L_{\odot}$ to 69 $L_{\odot}$ with a median of 1.3 $L_{\odot}$ has been reported \citep{b71}. Similarly, if the mass supply to the protostars from a depleting envelope gets reduced, their luminosities should also be reduced so that a T Tauri star should not be more luminous than protostars which are still in their embedded phase of collapse. The observations, however, clearly do not follow this prediction \citep{b52,b53}. 

A variety of initial conditions are present in star forming regions \citep{b31}. Observations have also revealed a considerable diversity in the morphology of young and developed star systems \citep{b32, b33}. When investigated using numerical models, this diversity in the morphology of the systems has been found to be highly dependent on the initial conditions prevailing in star forming gas
 \citep{b34, b35, b36, b37, b38}. The great number of physical parameters involved in setting the initial conditions in molecular cloud cores makes the phenomenon of star formation interesting as well as complex enough to explore in every detail. In addition to physical processes like episodic accretion onto young embedded protostars \citep{b2}, many other exciting features of star formation exist, such as protostellar jets \citep{b3, b4} observed during the accretion phase of protostars \citep{b5} and in the absence of efficient angular momentum transport mechanisms \citep{b59,b60}. The fundamental cause of episodic accretion is still unclear \citep{b39}.

Many authors from \citet{b19} to \citet{b20} have suggested the possibility of accretion via bursts that may last for relatively short durations within the range of timescales spanning a few tens to a hundred years. This was recently confirmed by performing zoom-simulations with the adaptive mesh-refinement code RAMSES \citep{b74}. The possibility of episodic mass inflow has also been found to be important in another phase of protostellar evolution known as the episodic molecular outflow phase \citep{b21}. 

\citet{b43} conducted a numerical study in which long-term simulations of the main accretion phase of star formation were carried out. They observed that spiral density waves are an important factor contributing to the strong time variability of the mass accretion rate onto protostars. They found that the time variability of the accretion rate is lower in clouds with lower initial rotational energy. Their results suggest a dependence of the mass accretion rate on the initial rate of rotation of a molecular core. Our simulations complement these previous findings by focusing on additional parameters that may also dictate the history of the mass accretion rates.

A large ensemble of simulations of collapsing and fragmenting cores has been performed by \citet{b300} to understand the observed characteristics of star formation in the Ophiuchus cloud. In this work they explored three types of models (i.e. no radiative feedback, continuous radiative feedback, and episodic radiative feedback) that affect fragmentation and the resulting protostellar masses. While their simulations without radiative feedback produced too many brown dwarfs, the ones with continuous radiative feedback yielded too few of them. However, with episodic radiative feedback both the peak of the protostellar mass distribution at $ \sim 0.2 M_{\odot}$ and the ratio of H-burning stars to brown dwarfs was found to be consistent with the observations.

While analyzing the evolving protostars in simulations we focus on how episodic accretion plays a possible role in the formation of multiple low-mass protostellar systems exhibiting vigorous mass accretion rates that may introduce even more complexity to the phenomenon of competitive mass accretion. To be able to follow collapsing systems on a sufficiently long timescale, we use a sink particle scheme combined with the well-known SPH method. The sink particle scheme replaces self-gravitating high density peaks within the collapsing gas with particles that only feel the gravitational influence of other mass-carrying particles in the simulation. Once a sink is created the material is allowed to flow smoothly towards the sink and can be accreted by the sink particle according to a set of rules that mimic accretion onto protostars as closely as possible \citep{b51}. In a similar way, the sink particle scheme can also be used to simulate accretion onto binaries and multiple stellar systems. In this way the sink particles act as sub-grid models for protostars without having to calculate their internal evolution in any detail. This strategy allows us to follow the detailed collapse of molecular cloud cores without suffering from the limitations imposed by ever decreasing time steps associated with growing densities within protostars as explained by \citet{b51}. The sink particle approach has been implemented in both Eulerian and Lagrangian numerical codes based on the AMR and SPH methods \citep{b26,b40,b64}. The sink particle scheme also enables us to analyze the physical processes going on in near protostellar clumps and allows us to study possible episodic accretion onto the sink particles along with their effects on nearby gas and other condensations in their vicinity.

The previous paragraphs contain the motivation and background for the numerical study presented in this paper. We have attempted to model the initial stages of the formation of systems composed of very low mass stars evolving within a common envelope of gas. Our models start with molecular cloud cores with varying initial temperatures and associated sound crossing times. These cores are also seeded with non-axisymmetric density perturbations of varying strength to mimic the presence of large scale turbulence on the level of molecular cloud cores. We find that the models give rise to a variety of outcomes. Our results include, for example, both multiple and single protostellar systems each consisting of very low-mass (VLM) protostars. Among the multiple protostellar systems we find both close and wide binaries. Some of these binary systems exhibit signs of episodic accretion. In addition to VLM binaries we also find VLM triplets which include binary systems as one of their components.

The plan of the paper is as follows. Section 2 discusses the computational scheme. In section 3 we describe the initial conditions and the setup for our models. Section 4 discusses the equation of state used in our models. Section 5 briefly explains the treatment of self-gravitating clumps and a reference to the scheme adopted to compute binary properties. Section 6 then provides a detailed account of our results and finally we present our conclusions in section 7.

\section{Computational Scheme}
The numerical models presented in this paper are based on the particle simulation method known as smoothed particle hydrodynamics (SPH). We use the computer code GRADSPH developed by \citet{b22}. GRADSPH is a fully three-dimensional SPH code which combines hydrodynamics with self-gravity and has been specially designed to study self-gravitating astrophysical systems such as molecular clouds. 
The numerical scheme implemented in GRADSPH implements the long range gravitational interactions within the fluid by using a tree-code gravity (TCG) scheme combined with the variable gravitational softening length method \citep{b23}, whereas the short range hydrodynamical interactions are solved 
using a variable smoothing length formalism.  The code uses artificial viscosity to treat shock waves. 

In SPH, the density $\rho_{i}$ at the position $\vec{r}_{i}$ of each particle with mass $m_{i}$ is determined by summing the contributions from its neighbors using a weighting function 
$W\left(\vec{r}_{i}-\vec{r}_{j},h_{i}\right)$ with smoothing length $h_{i}$:

\begin{equation} \label{density}
\rho _{i} =\sum _{j} m_{j} W\left(\vec{r}_{i}-\vec{r}_{j},h_{i}\right).
\end{equation}

GRADSPH uses the standard cubic spline kernel with compact support within a smoothing sphere of size $2 h_{i}$ \citep{b23}.
The smoothing length $h_{i}$ itself is determined using the following 
relation: 

\begin{equation} \label{smoothinglength}
h_{i}=\eta \left(\frac{m_{i}}{\rho_{i}}\right)^{1/3}, \ \ \ \ \ 
\end{equation}

where \textit{$\eta$} is a dimensionless parameter which determines 
the size of the smoothing length of the SPH particle given its mass and density. 
This relation is derived by requiring that a fixed mass, or equivalently a fixed number of neighbors,  must be contained inside the smoothing sphere of each each particle:

\begin{equation} \label{fixedmass}
\frac{4\pi}{3}\left(2 h_{i}\right)^{3} \rho_{i}=m_{i} N_{opt}=constant. \ \ \ 
\end{equation}

Here N$_{opt}$ denotes the number of neighbors inside the smoothing sphere, which we set equal to
50 for the 3D simulations reported in this paper. 
Substituting Eq. \ref{smoothinglength} into Eq. \ref{fixedmass} allows us to determine $\eta$:

\begin{equation} \label{eta}
\eta=\frac{1}{8}\left(\frac{3 N_{opt}}{4 \pi}\right)^{1/3}.
\end{equation}

Since Eq. \ref{density} and Eq. \ref{smoothinglength} depend on each other, we solve the two equations iteratively at each time step and for each particle.

The evolution of the system of particles is computed using the second-order predict-evaluate-correct(PEC) scheme implemented in GRADSPH, which integrates the SPH form of the equations of hydrodynamics with individual time steps for each particle. For more details and a derivation of the system of SPH equations we refer to \citet{b22}.

\section{Initial conditions and setup}

In the present study we use a modified version of the Boss and Bodenheimer collapse test with initial conditions described in \citet{b24}. The overall setup of the initial conditions is identical to the cloud core models that we used previously while investigating the thermal response of collapsing molecular cores \citep{b25}. We stick to m = 2 non-axisymmetric density perturbations with a fixed amplitude A that varies from model to model. We introduce identical amplitudes of the azimuthal density perturbation in each hemisphere of the uniform density core models. This perturbation is designed to approximately include the effect of large scale turbulence on the scale of giant molecular clouds, which can induce asymmetric density distributions in embedded molecular cores that are on the verge of gravitational collapse. For this purpose we use the following form of the density perturbation

\begin{equation} \label{perturb}
\rho=\rho_{0}\left\{1+A\sin\left(m\varphi{}\right)\right\}\ \ \ \ \ \ 
\end{equation}

where $\varphi$ is the azimuthal angle in spherical coordinates (r,$\varphi$, z).

Our solar mass molecular cloud core models with an initial temperature of 8 K have dimensionless ratios of Jeans mass to initial core mass, thermal energy to magnitude of gravitational energy, and rotational energy to the magnitude of the gravitational potential energy equal to 0.143, 0.264, and 0.0785, respectively. Models with other initial thermal states have corresponding values for these ratios.
While we explore a certain range of parameters, we also emphasize that statistically significant trends can only be derived after exploring a more extensive range of possible realizations for the initial conditions. While our results may provide hints towards possible trends or general phenomena, we do not intend here to derive strict relations. We emphasize that the current paper thus provides only a first step, and a much larger range of initial conditions needs to be explored in the future. 

\section{Equation of state}

The chemical composition of the clouds is similar our previously studied models \citep{b25}. We assume the gas to be a mixture of hydrogen and helium with mean molecular weight $\mu$ = 2.35. The initial temperature $T_{0}$ is chosen as 8 K for the models labeled M1, M2, M3, and M4 presented in this paper. The corresponding sound speed 
is $c_{s}$ = 167.62 m/s and the initial free fall time is $t_{ff}$ = 1.07 x 10$^{12} s$ or 3.4 x 10$^{4}$ yr. Furthermore, the sound speed takes different values for models M5 and M6 as we vary the initial thermal state of the molecular core by setting the initial temperature to 10 K and 12 K, respectively. We adopt a barotropic equation of state of the form 

\begin{equation} \label{EOS}
P=\rho c_{0}^{2}\left[1+\left(\frac{\rho}{\rho_{crit}}\right)^{\gamma-1}\right],
\end{equation}

in which $\gamma$ = 5/3. This approximate equation of state describes the gradual transition from isothermal to adiabatic behaviour of the gas during gravitational collapse. For the critical density \textit{$\rho$}$_{crit}$ we use a value of 10$^{-13}$ g/cm$^{3}$ which is slightly higher than the value of 10$^{-14}$ g/cm$^{3}$. 
An overview of the models is provided in table 1. The table contains the properties of the initial conditions and relevant information on the final outcome of their evolution. 
By adopting various values for the azimuthal density perturbation amplitudes while keeping the rest of the initial parameters identical for models M1 to M4, and testing different initial thermal states for models M5 and M6, we aim to investigate the role of primary/secondary fragmentation in the formation of VLM stars. By primary fragmentation we mean the breakup of a cloud into initial fragments as the direct result of the imposed perturbation, whereas secondary fragmentation mainly takes place in later stages during the course of the further evolution of the system. 
This strategy gives us the opportunity to analyze numerically the properties of weak or strong primary-fragmentation. Primary fragmentation can control the mass accretion process onto secondary condensations and hence can dictate the eventual outcome of the collapse. While varying the perturbation strengths and the initial thermal states, we study the mass ratio, the eccentricity, the binary separation and the type of binary companion of the resulting binary systems. We also look at the rate of accretion of individual fragments and the number of primary/secondary fragments in the embedded phase of core collapse.

We note again that the relations which result from this investigation need to be regarded with a grain of salt, and should be tested in future studies employing a larger range of initial conditions.

\section{Treatment of self-gravitating clumps}

In order to deal with self-gravitating clumps in our simulations, 
we took advantage of the recently proposed improved sink particle algorithm for SPH calculations described by \citet{b26}. We introduce sinks into the version of GRADSPH used in this paper following the algorithm NEWSINKS by \citet[][see sec. 2]{b26} in which an SPH particle can be turned into a sink with accretion radius $r_{sink}$ once its density exceeds a given threshold $\rho_{sink}$. For the present simulations we adopt the values $\rho_{sink}$=10$^{-10}$ g/cm$^{3}$ and $r_{sink}$=1.0 au, respectively. 
The criteria for sink particle creation and accretion of SPH particles within the accretion radius follow sections 2.2 and 2.3 in \citet{b26}, respectively. Because the mechanism for angular momentum transfer in accretion disks is still uncertain, we decided not to implement angular momentum feedback from sink particles.

For computing binary properties of sinks emerging in our simulations we take advantage of a scheme discussed by \citet{b61} in their section 2.5.

\section{Results and Discussion}

We now state the results of our simulations and discuss them systematically. 

\subsection{Dependence on perturbation amplitude}

In this paper we intend to investigate how the first fragments that form in our simulations affect the later secondary fragmentation processes within disks orbiting protostars. The fragments which appear as a result of the initial perturbation with azimuthal dependence depending on the mode m and the amplitude A are designated as primary fragments. On the other hand, clumps that emerge because of gravitational instabilities arising in evolving disks within the core at a later stage and which are not directly connected to the initial perturbation are regarded as secondary fragments.   
   
The evolution of model M1 is illustrated in Figure 1. The plots are face-on column density plots along the rotation axis at 6 successive times during the evolution of the model. Two primary fragments (sink 1 and sink 2) emerge at 42.40 kyr and 42.51 kyr, respectively. As the system evolves further, it gives birth to 4 additional secondary fragments (sink 3, sink 4, sink 5 and sink 6) which appear at 42.58 kyr, 42.59 kyr, 42.94 kyr and 42.99 kyr, respectively. The small amplitude of the initial density perturbation A = 0.005 in this model does not seem to favor the primary fragments in terms of mass accumulation and all primary and secondary fragments eventually accrete enough material from the envelope to become VLM stars. Panel b shows that some of the protostars form close binary systems as depicted by the zoomed in part of the panel. There is also a triplet system indicated by an arrow in panel f which includes a close binary system.  

Although we have not followed the dynamics of this system further in time (primarily due to computational constraints such as too small time steps), there are indications that the development of such an initially compact triplet could result in the formation of a wide binary system (provided that there is not much gas available for dynamical friction to shrink the binary separation) on a time scale of millions of years as one component is migrated to a distant orbit at the cost of shrinking the orbits of the other two components in the system \citep[see e.g.][]{b62,b63}. The presence of a relatively small initial perturbation in model M1 seems to support the formation of a relatively large number of sinks. Furthermore, the triplet system which is formed during the evolution of this model only consists of secondary fragments. By the time the simulation ends at t = 3.00 $t_{ff}$, in addition to an unbound primary, the binary (primary-secondary) and the triplet (secondaries) are left as gravitationally bound systems embedded within the collapsing gas with binary separations of 28.61 au and 25.00 au and mass ratios of 0.752 and 0.93, respectively for the two bound systems as summarized in table 1 and table 4.

Figure 2 shows the evolution of model M2. In this model the strength of the initial density perturbation increases by a factor of 10 compared to model M1. The two primary fragments (sink 1 and sink 2) emerge in this model at 42.70 kyr and 42.76 kyr, respectively. Further gravitational collapse of the core produces additional secondary fragments (sink 3, sink 4, sink 5) at 43.13 kyr, 43.94 kyr, and 46.47 kyr, respectively.

In model M2 we find a binary system composed of sink 2 and sink 4 i.e. a (primary-secondary) combination. The binary separation at this stage is 46.48 au and the eccentricity of the system is 0.22 (see table 4).

Figure 3 presents the evolution of model M3 in which the level of the initial density perturbation is further increased by a factor of 2. In this case the primary fragments are clearly able to accrete a significant fraction of the gas present in the collapsing core. The two primary fragments (sink 1 and sink 2) appear at 40.64 kyr and 44.46 kyr, respectively. Thereafter, 3 more secondary fragments are formed (sink 3, sink 4, and sink 5) at core evolution times 44.87 kyr, 46.78 kyr, and 52.25 kyr, respectively. In this model one could expect to observe an enhanced mass contrast among the fragments which could lead to the formation of VLM stars, brown dwarfs, or even planemos but it does not seem that such a trend prevails and decides the outcome of the core collapse. The evolution of this model yields 4 VLM stars which form two close binaries with binary separations of 30.53 au and 38.47 au, respectively. In addition to these close binaries there is also a single protostar which is unbound with respect to the nearby binary systems.   
Compared to model M1, the total number of fragments roughly remains the same. At the end of the simulation at t = 2.998 $t_{ff}$, the system again contains a single VLM star and a couple of VLM binary systems which are a combination of primary-secondary and secondary-secondary fragments (see table 4).

\begin{figure*} \label{fig:1}
\centering
\includegraphics[angle=0,scale=0.5]{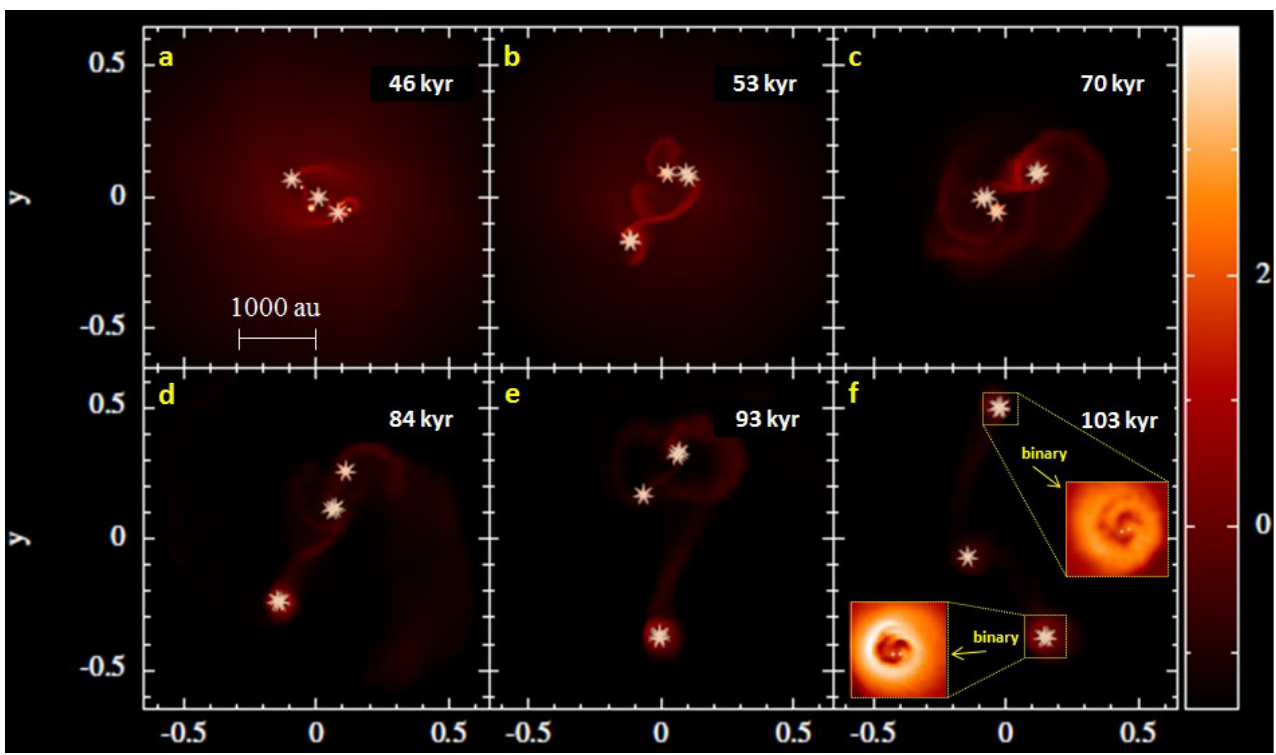}
\caption{Results of the simulation for model M1. The physical size of each individual panel is 4333 x 4333 au. The plots show face-on views of the column density integrated along the rotation axis(z) at 6 successive times during the evolution of the model. The colour bar on the right shows log ($\Sigma{}$) in dimensionless units. Each calculation was performed with 250025 SPH particles.}
\end{figure*}

\begin{figure*} \label{fig:2}
    \centering
    \includegraphics[angle=0,scale=0.5]{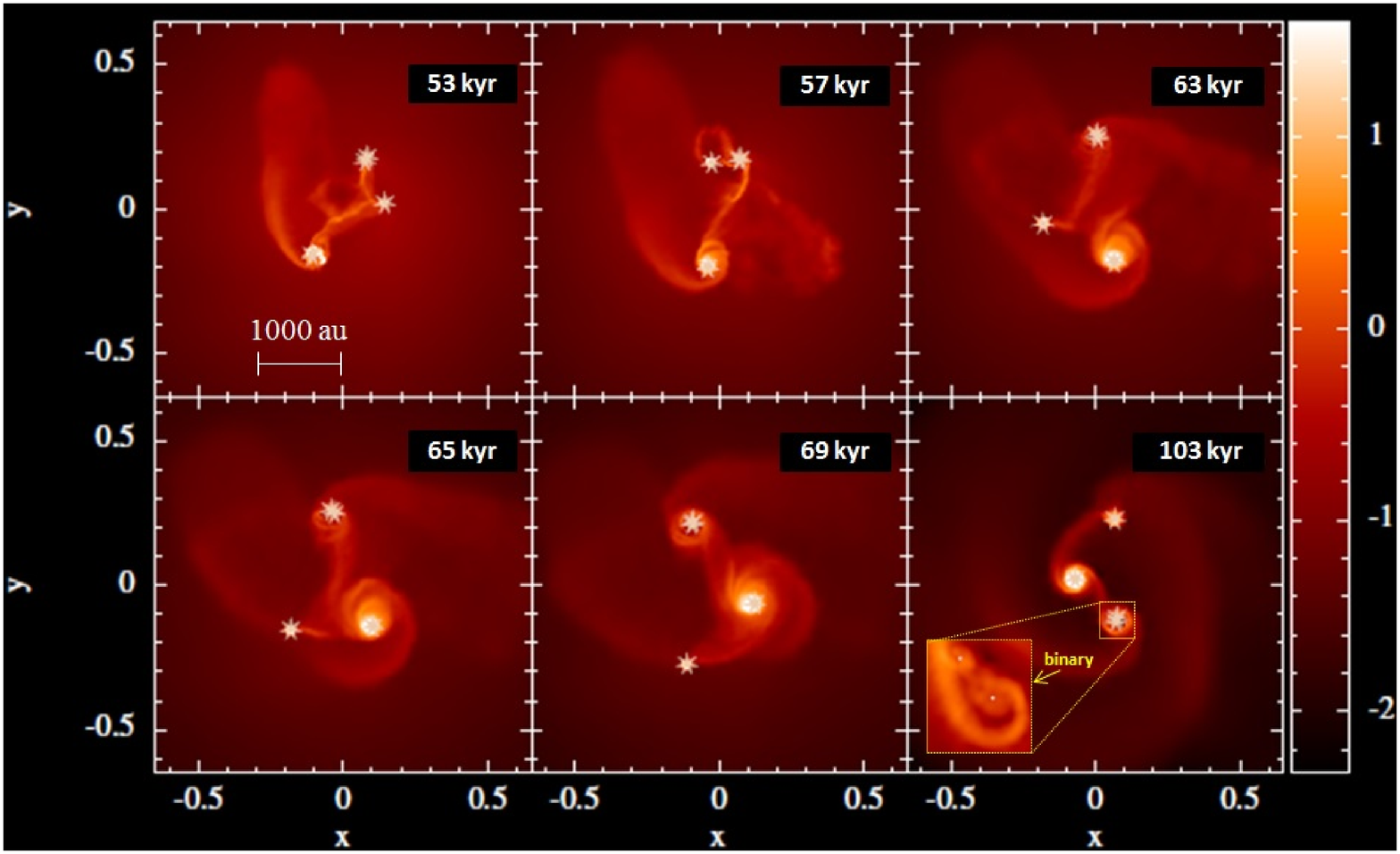}
    \caption{Results of the simulation for model M2. The physical size of each individual panel is 4333 x 4333 au. The plots show face-on views of the column density integrated along the rotation axis(z) at 6 successive times during the evolution of the model. The colour bar on the right shows log ($\Sigma{}$) in dimensionless units. Each calculation was performed with 250025 SPH particles.}
  \end{figure*}
  
\begin{figure*} \label{fig:3}
    \centering
    \includegraphics[angle=0,scale=0.5]{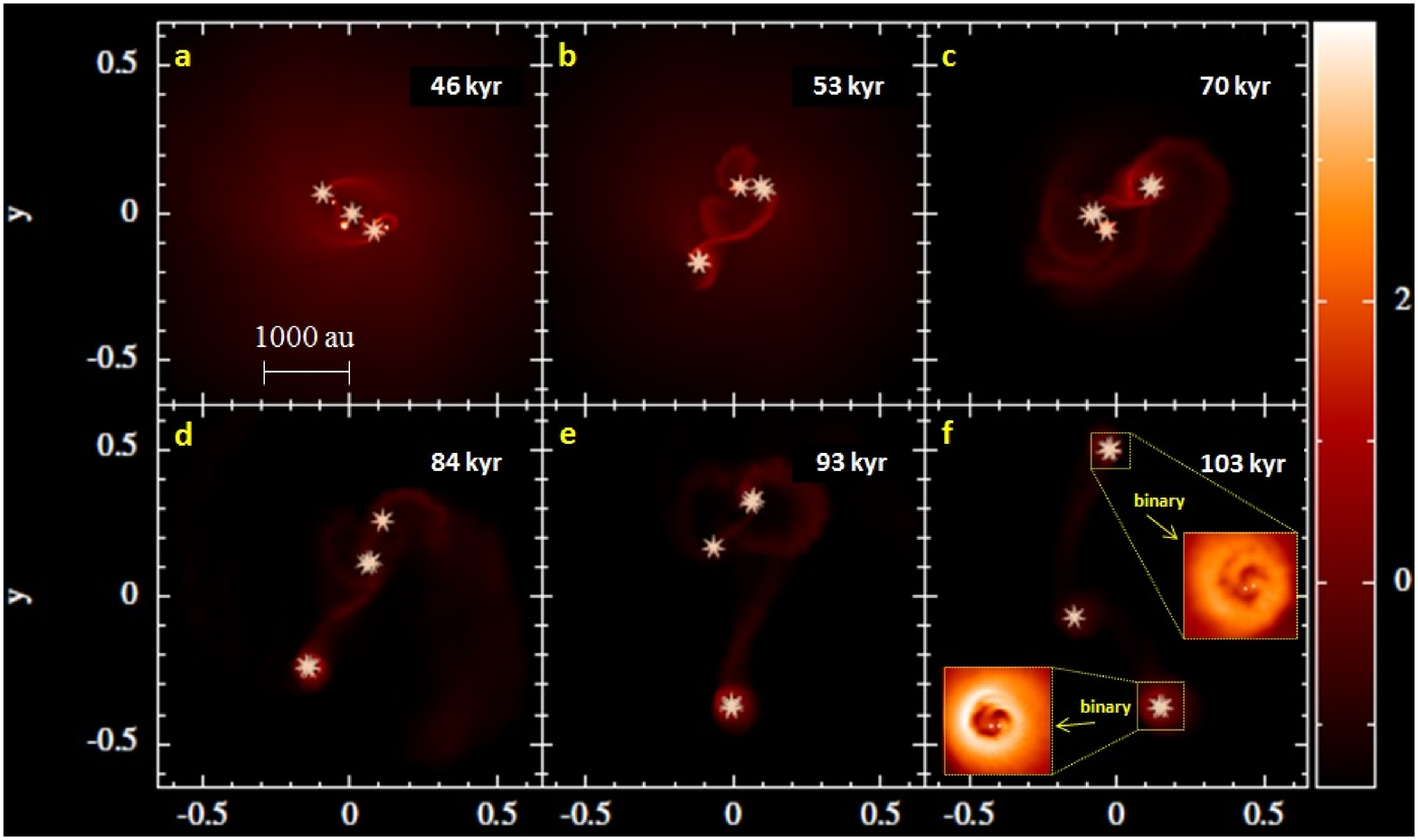}
    \caption{Results of the simulation for model M3. The physical size of each individual panel is 4333 x 4333 au. The plots show face-on views of the column density integrated along the rotation axis(z) at 6 successive times during the evolution of the model. The colour bar on the right shows log ($\Sigma{}$) in dimensionless units. Each calculation was performed with 250025 SPH particles.}
  \end{figure*}

\subsection{Dependence on initial temperature}

We now turn to the effect of increasing the temperature of the core. In model M5, we raised the initial temperature by 2 K to study the effect of a shorter sound crossing time and a higher initial thermal state on the collapse. The remainder of the model parameters were identical to model M2. The results are reported in Figure 5. 
The primary fragments (sink 1 and sink 2) are formed at 45.69 kyr and 45.96 kyr, respectively. Two secondary fragments emerge as sink 3 and sink 4 at 46.09 kyr and 63.37 kyr. Compared to model M2, model M5 takes a longer time to give birth to the last fragment in the core. Although the density perturbation is relatively modest, the change in initial temperature greatly alters the outcome of this model. 
At the end of the simulation for model M5 at t = 7.307 $t_{ff}$, two binary systems with components of identical nature have appeared. One is a very close VLM binary composed of two secondary fragments with binary separation and mass ratio of 28.180 au and 0.496, respectively, while the other is a relatively wide VLM binary including two primary fragments and with binary separation and mass ratio of 231.50 au and 0.723, respectively. The binding energy of the secondary-secondary binary will be much greater than for the primary-primary binary since both have a similar mass and the primary-primary has a much wider separation.

Figure 6 illustrates the development of the last model M6 where the initial temperature has been further increased by 2 K compared to model M5. As expected, the increase in the thermal energy content of the collapsing molecular core delays the fragmentation process. This is directly related to how the temperature change affects the Jeans mass at the density of fragmentation \citep{b67}. The Jeans mass at density $\rho$ and temperature T can be expressed as 

\begin{equation} \label{Jeans Mass}
M_{J}=1.86 \left(\frac{RT}{G}\right) ^{3/2} {\rho}^{-1/2}.
\end{equation}

Eq. (\ref{Jeans Mass}) shows that for a given density, the Jeans mass takes on larger values at higher temperatures so that for any potential fragment, it therefore takes a relatively longer time to reach the threshold above which gravitational collapse occurs.

The two primary fragments emerge within the collapsing cloud as sink 1 and sink 2 at 51.78 kyr and 56.01 kyr, respectively. This is followed by the appearance of a third and last fragment (sink 3) 57.05 kyr after the start of the collapse. The presence of only 3 fragments after 0.162 Myr of evolution clearly shows that gravity is partly balanced by the thermal pressure exerted by the gas. Model M6 evolves into a protobinary system of the primary-primary type with binary separation and mass ratio equal to 110.50 au and 0.858, respectively, and there is also one single VLM star. Although we did not run a statistically significant number of simulations to be able to perform a thorough analysis of the properties of the binaries which are formed in our models, we find interesting differences between the two sets of models (M1, M3, and M4) and (M2, M5, and M6). On the one hand, all of the binaries formed in the first set are composed of a combination of primary and secondary fragments. On the other hand, all of the binaries formed in the second set of models are either a combination of two primary fragments or two secondary fragments (see table 4 for details).
Future work will be needed to explore in more detail if such a relation can indeed be confirmed.

\begin{table*} \label{tbl-1}
\small 

\begin{flushleft}
\centering
\caption{Summary of the initial physical parameters and the final outcome of the models discussed in this paper. The initial radius, mass, average density and rate of rotation for each model are given by the constant values 5 x 10$^{16}$ cm, 1 $M_{\odot}$, 3.8 x 10$^{-18}$ g/cm$^{3}$ and 5.0 x 10$^{-13}$ rad/s, respectively. }
\begin{tabular}{cccccc}
\hline
\hline
Model & A & $T_{0}$(K) & Final outcome \\
\hline
M1  & 0.005   &  8  & single bound VLM triplet, two bound VLM binaries, single bound VLM star \\
M2  & 0.05    &  8  & one bound VLM binary, two single bound VLM stars \\
M3  & 0.1	  &  8  & two bound VLM binaries, single bound VLM star \\
M4  & 0.2     &  8  & single bound triplet, single bound VLM star \\
M5  & 0.05    &  10 & two bound VLM binaries \\
M6  & 0.05    &  12 & single bound VLM binary, Single bound VLM star \\

\hline
\end{tabular}
\end{flushleft}
\end{table*}

\begin{table*} \label{tbl-2}
\small 

\centering
\caption{Summary of the final evolution time $t_{f}$, the total number of fragments $N_{tot}$, the masses of the fragments (in the order as they appear in the simulation), and the number of *bound* objects (VLM stars) $N_{bound}$ at the end of the simulation. }
  \begin{tabular}{cccccc}
\hline
\hline
Model & $t_{f}$ (kyr) & $N_{tot}$ & Fragment masses ($M_{\odot}$) & $N_{bound}$ \\
\hline 
M1  & 103  & 6  &  0.1917, 0.1882, 0.1441, 0.1184, 0.1153, 0.0087     &  5 \\
M2  & 103  & 5  &  0.1808, 0.1628, 0.1107, 0.1663, 0.1474             &  4 \\
M3  & 103  & 5  &  0.2252, 0.2045, 0.1739, 0.1190, 0.1124             &  4 \\
M4  & 185  & 4  &  0.2470, 0.2962, 0.1868, 0.1030                     &  4 \\
M5  & 250  & 4  &  0.2555, 0.1850, 0.2745, 0.1364                     &  4 \\
M6  & 163  & 3  &  0.2347, 0.2013, 0.2488                             &  3 \\

\hline
\end{tabular}
\end{table*}

\begin{table*} \label{tbl-3}
\centering
 \caption{Summary of the maximum evolution time in units of the initial free fall time $t_{ff}$, and the final ratio of the envelope mass to the initial mass of cloud core for models M1$\rightarrow{}$M6.}
 \begin{tabular}{ccccccc}
\hline
\hline
 Model & M1 & M2 & M3 & M4 & M5 & M6 \\ 
\hline 
 $t_{ff}$           & 3.000 & 2.998 & 2.998 & 5.401 & 7.307 & 4.764  \\
 $M_{env}/M_{core}$ & 0.154 & 0.232 & 0.165 & 0.167 & 0.149 & 0.316  \\
\hline
\end{tabular}
\end{table*}

\begin{table*} \label{tbl-4}
\centering
 \caption{Summary of the models, type of the fragments which are components of binaries, masses of the components, mass ratio (q), semi-major axis (a), eccentricity (e), and binary separation (d) at the end of each run.} 
 \begin{tabular} {ccccccc}
\hline
\hline
 Model & component types & component masses ($M_{\odot}$) & q & a (au) &  e  &  d(au) \\
 \hline
M1 & primary 1 - secondary 1   & 0.1917, 0.1441 & 0.752 & 28.43  & 0.712 & 28.61 \\
M1 & secondary 2 - secondary 3 & 0.1184, 0.1153 & 0.973 & 45.43  & 0.491 & 25.00  \\
M2 & primary 2 - secondary 2   & 0.1628, 0.1663 & 0.979 & 43.78  & 0.220 & 46.48  \\
M3 & primary 1 - secondary 3   & 0.2252, 0.1124 & 0.499 & 37.09  & 0.273 & 30.53  \\
M3 & secondary 1 - secondary 2 & 0.1739, 0.1190 & 0.684 & 40.44  & 0.056 & 38.47  \\
M4 & primary 1 - secondary 1   & 0.2470, 0.1868 & 0.756 & 53.90  & 0.170 & 56.30   \\
M5 & primary 1 - primary 2     & 0.2555, 0.1850 & 0.723 & 366.80 & 0.960 & 231.50  \\
M5 & secondary 1 - secondary 2 & 0.2745, 0.1364 & 0.496 & 22.13  & 0.273 & 28.18  \\
M6 & primary 1 - primary 2     & 0.2347, 0.2013 & 0.858 & 128.40 & 0.366 & 110.50  \\
\hline
\end{tabular}
\end{table*}
 
\begin{figure*} \label{fig:4}
    \centering
    \includegraphics[angle=0,scale=0.5]{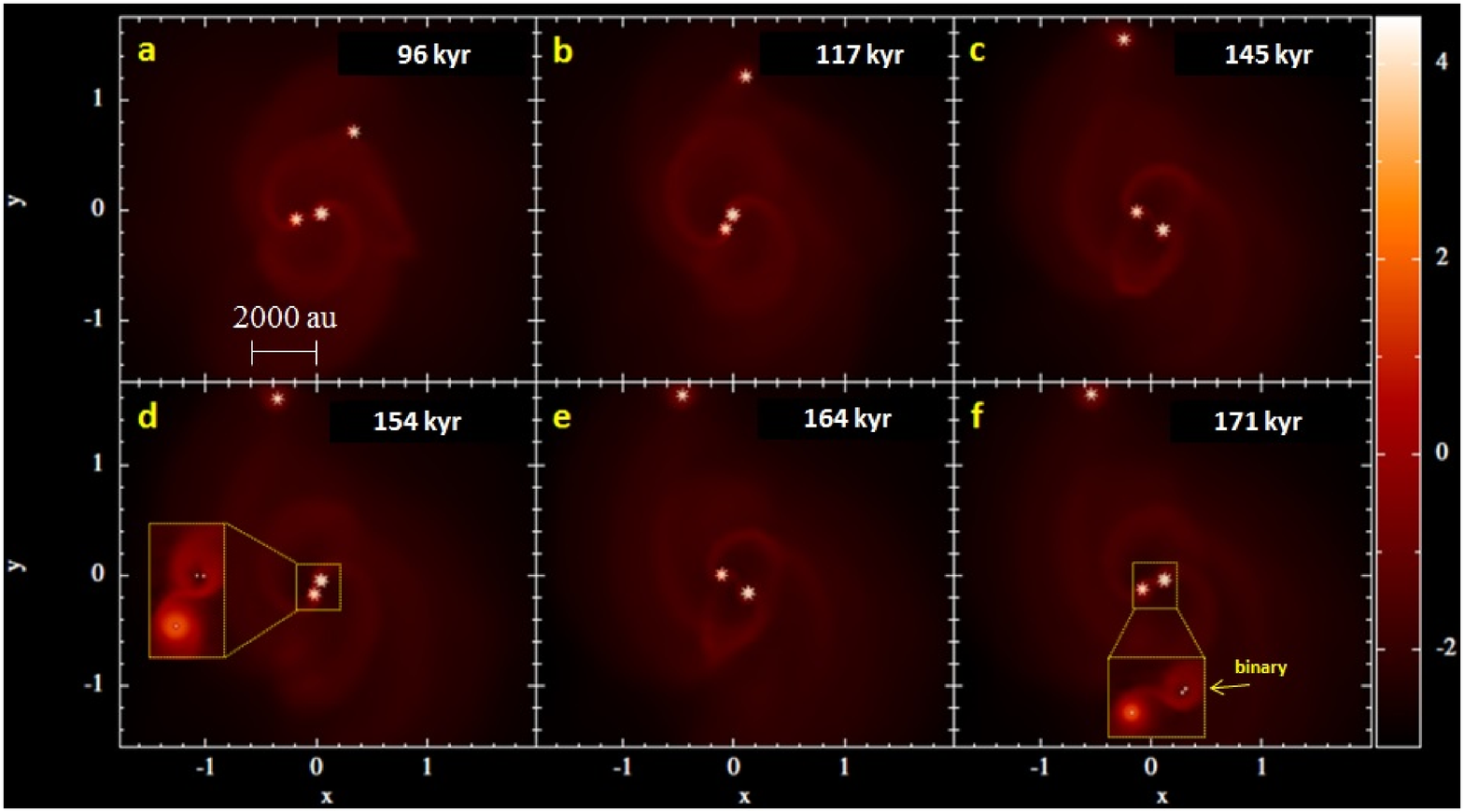}
    \caption{Results of the simulation for model M4. The physical size of each individual panel is 12667 x 12667 au. The plots show face-on views of the column density integrated along the rotation axis(z) at 6 successive times during the evolution of the model. The colour bar on the right shows log ($\Sigma{}$) in dimensionless units. Each calculation was performed with 250025 SPH particles.}
  \end{figure*}
  
\begin{figure*} \label{fig:5}
    \centering
    \includegraphics[angle=0,scale=0.5]{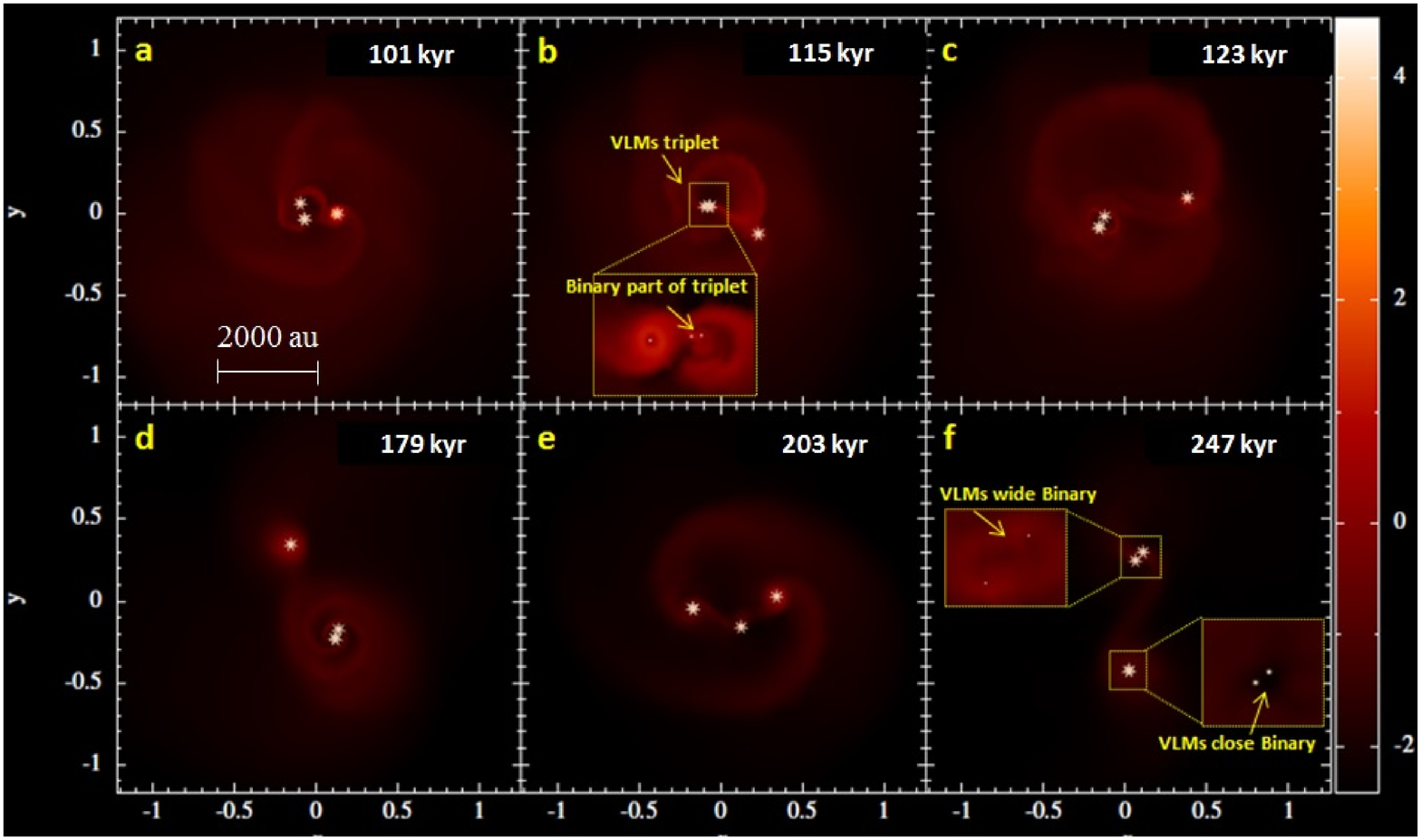}
    \caption{Results of the simulation for model M5. The physical size of each individual panel is 12667 x 12667 au. The plots show face-on views of the column density integrated along the rotation axis(z) at 6 successive times during the evolution of the model. The colour bar on the right shows log ($\Sigma{}$) in dimensionless units. Each calculation was performed with 250025 SPH particles.}
  \end{figure*}
  
\begin{figure*} \label{fig:6}
    \centering
    \includegraphics[angle=0,scale=0.5]{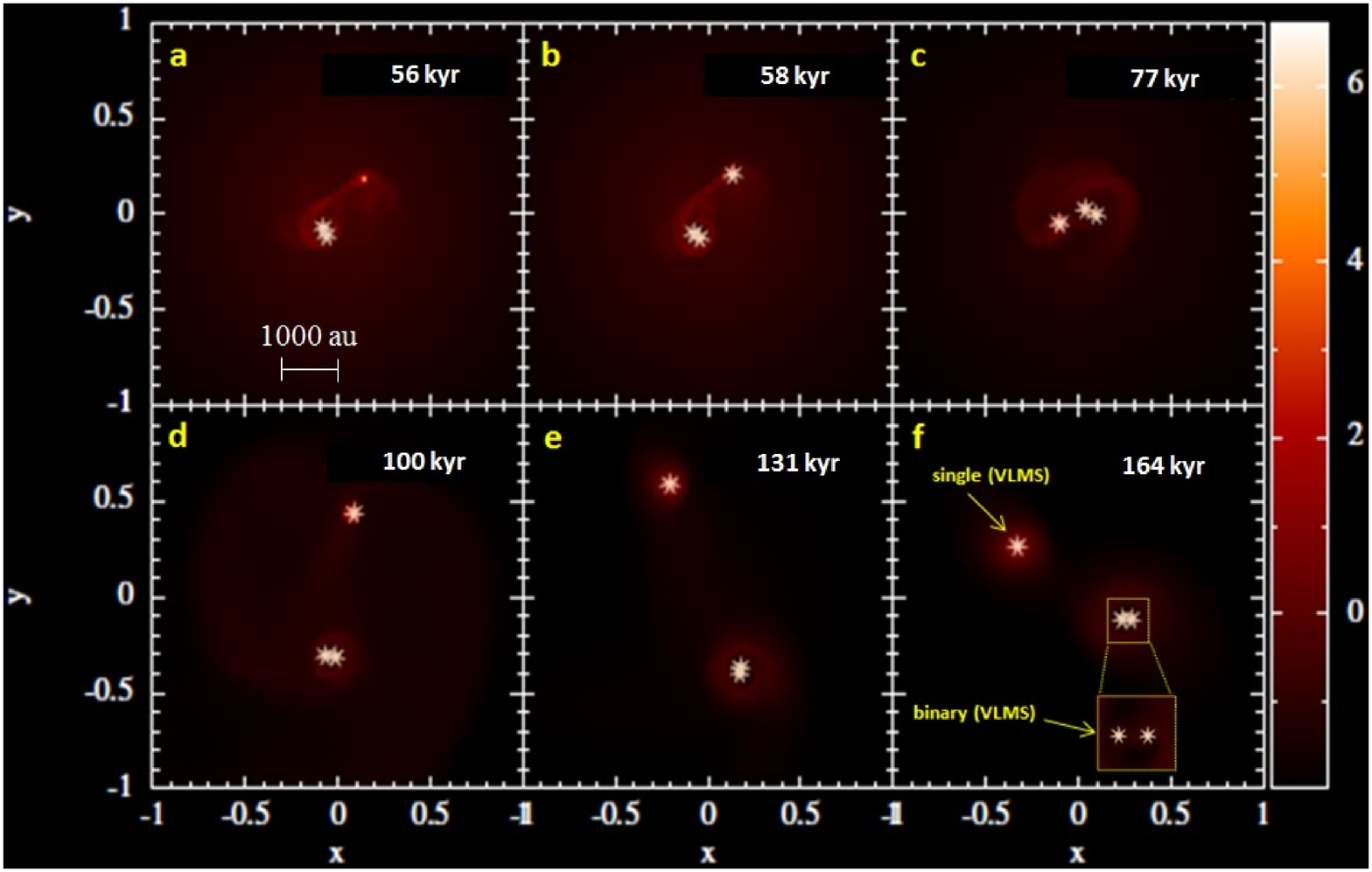}
    \caption{Results of the simulation for model M6. The physical size of each individual panel is 6666 x 6666 au. The plots shows face-on views of the column density integrated along the rotation axis(z) at 6 successive times during the evolution of the model. The colour bar on the right shows log ($\Sigma{}$) in dimensionless units. Each calculation was performed with 250025 SPH particles.}
  \end{figure*}
  
  
       \begin{figure*} \label{fig:7}
    \centering
    \includegraphics[angle=0,scale=0.75]{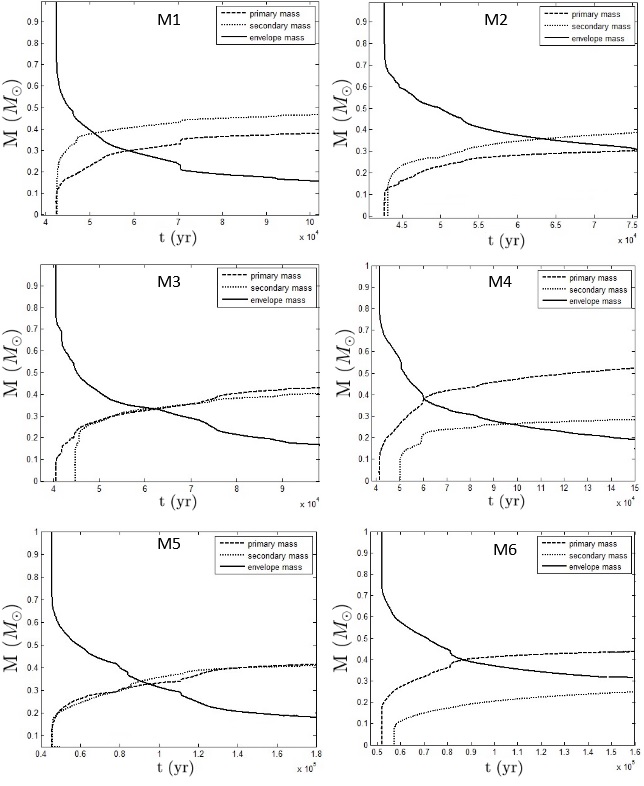}
    \caption{Time evolution of the total mass contained in primary fragments (dashed lines), secondary fragments(dotted lines) as well as the remaining mass in the envelope of the core (full lines). The curves show the entire evolution for models M1$\rightarrow{}$M6. The unit of mass is one solar mass and the time is expressed in years.}
     \end{figure*}

 \begin{figure*} \label{fig:8}
    \centering
    \includegraphics[angle=0,scale=0.75]{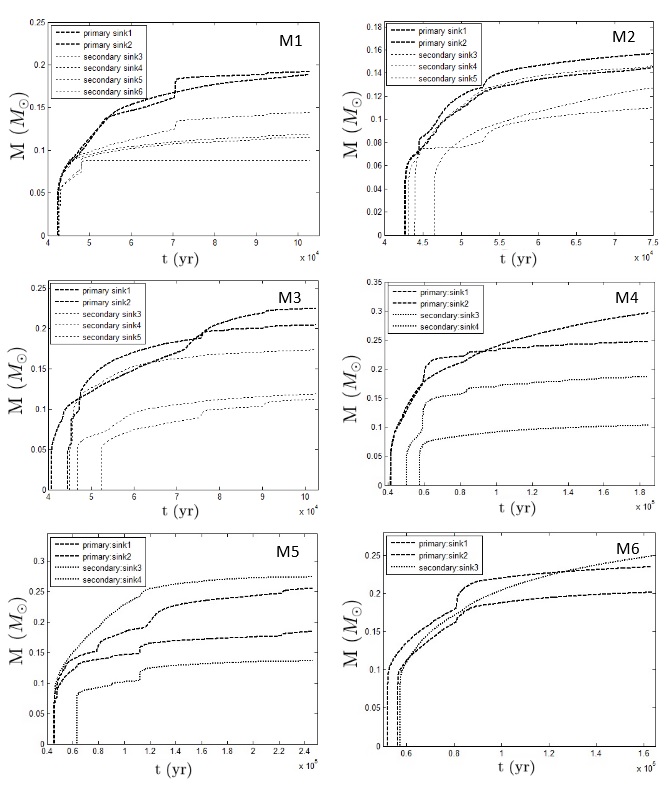}
    \caption{Evolution of the mass of the individual protostellar embryos for models M1$\rightarrow{}$M6. The unit of mass is solar mass and the time is in years. Each protostar corresponds to a sink particle in the simulation. Primary fragments are shown as dashed lines while secondary fragments correspond to dotted lines, respectively. }
     \end{figure*}
     
\begin{figure*} \label{fig:9}
\centering
\includegraphics[angle=0,scale=0.42]{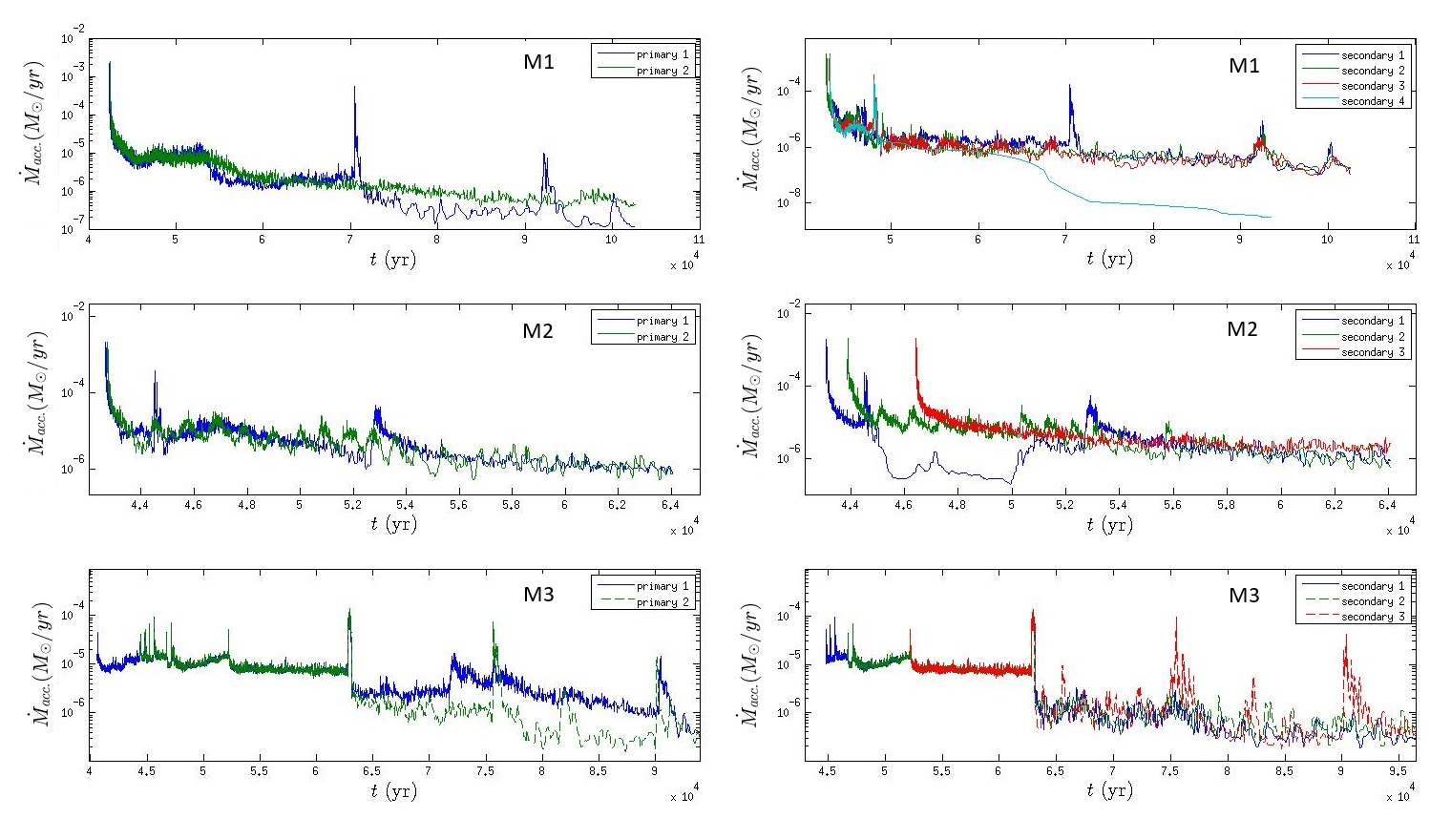}
\caption{Evolution of the mass accretion rate of the individual protostellar embryos during the evolution of models M1-M3. In model M1 the binares are (primary 1 \& secondary 1) and (secondary 2 \& secondary 3). In model M2 the binary is (primary 2 \& secondary 2). In model M3 the binaries are (primary 1 \& secondary 3), (secondary 1 \& secondary 2). The accretion rates for primary protostars and secondary protostars of each model are  shown on left and right panels; respectively. The accretion rate for each protostar is shown in units of log of solar mass per year. The time is indicated in years. The first spike for each appearing sink in all panels represents formation of a protostar which is followed by subsequent accretion bursts.}
\end{figure*} 

\begin{figure*} \label{fig:10}
\centering
\includegraphics[angle=0,scale=0.42]{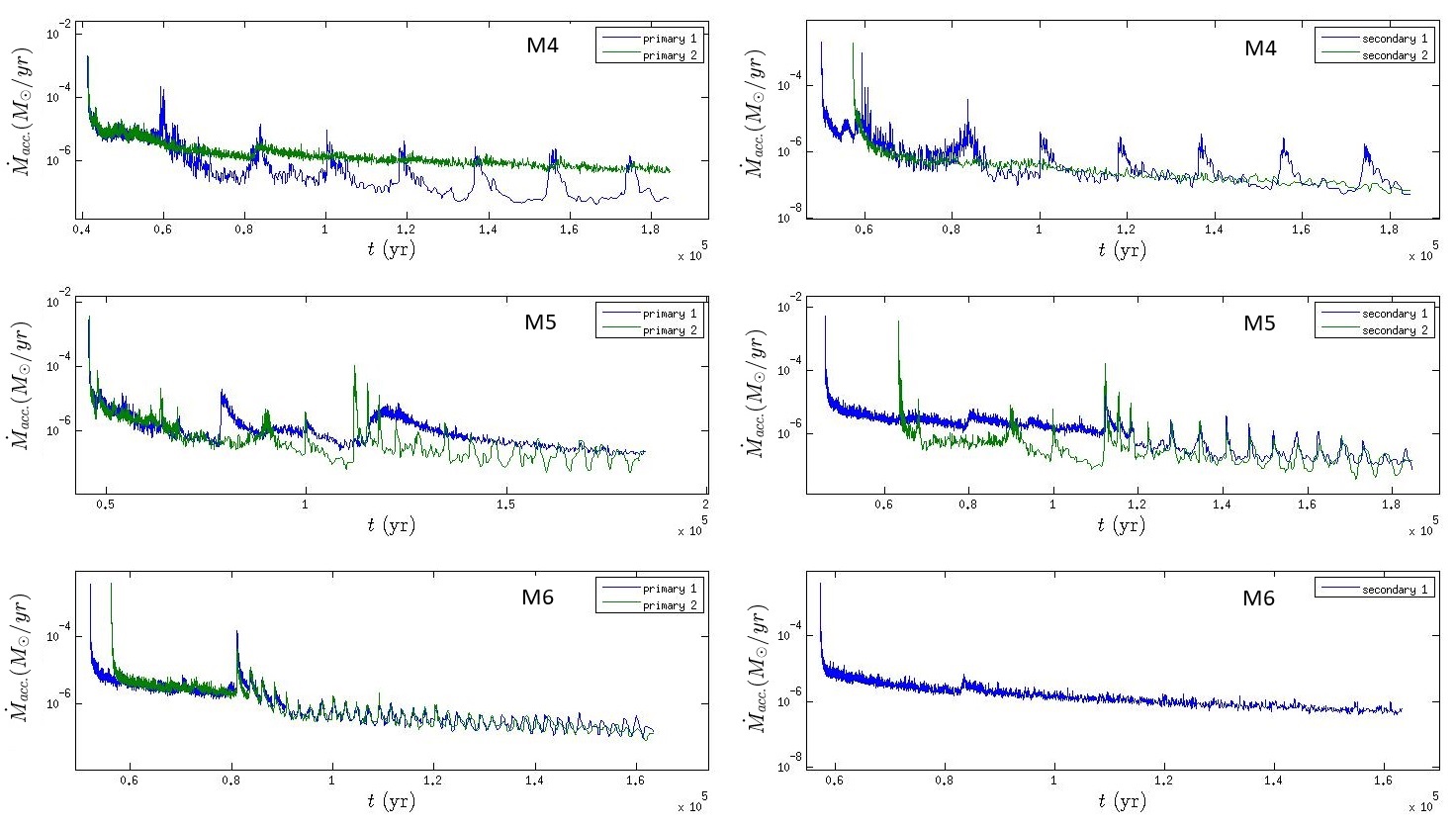}
\caption{Evolution of the mass accretion rate of the individual protostellar embryos during the evolution of models M4-M6. In model M4 the binary is (primary 1 \& secondary 1). In model M5 the binaries are (primary 1 \& primary 2) and (secondary 1 \& secondary 2). In model M6 the binary is (primary 1 \& primary 2). The accretion rates for primary protostars and secondary protostars of each model are  shown on left and right panels; respectively. The accretion rate for each protostar is shown in units of log of solar mass per year. The time is indicated in years. The first spike for each appearing sink in all panels represents formation of a protostar which is followed by subsequent accretion bursts.}
\end{figure*} 


\subsection{Evolution of clump masses and accretion rate}

In a recent study reported by \citet{b42}, the properties of prestellar cores that make core disks susceptible to fragmentation have been analyzed. These authors find that a higher initial core angular momentum and a higher core mass lead to more fragmentation, while higher levels of background radiation and magnetic fields moderate the tendency of the disk to fragment. On the other hand, the simulations reported here focus on how primary and secondary fragmentation takes place within the core disks and allow us to study the dependence of these phenomena on initial conditions with varying initial density perturbations as well as varying initial thermal states. 
With this goal in mind, Figure 7 shows the evolution of the total mass contained within the primary fragments (dashed lines) as well as the secondary fragments (dotted lines), along with the remaining mass in the envelope surrounding the fragments for models M1-M6 (full lines). Comparing models M1, M2, M3 and M4 shows that the total mass contained within secondary fragments formed within relatively evolved core disks is larger than the mass accreted by the primary fragments from the envelope when the amplitude of the initial density perturbation is weak. However, a gradual increase in the level of the initial density perturbation starts reversing this trend. In model M3, the mass contained within the primary fragments almost equals that within the secondary fragments while in model M4, which further increases the initial perturbation amplitude, the mass accumulated within the primary fragments has become clearly dominant. Figure 7 also illustrates the effect of increasing core temperatures on the mass accumulation history of the system when the perturbation level stays constant. This involves models M2, M5 and M6 which shows that a higher thermal state at the beginning of core collapse tends to suppress the accretion of mass by the secondary fragments within the core disk. So both stronger perturbations and higher initial temperatures seem to prefer primary fragmentation over secondary fragmentation in terms of their accumulation of mass, so that the primary fragments accrete more material from the core envelope than the secondary fragments.

Figures 8 shows the evolution of the mass of {\em individual} fragments within the cloud and serves to investigate the mass accretion history of individual primary and secondary fragments as well as the mass contrast between the different types of fragments. Primary fragments are indicated with dashed lines and secondary fragments with dotted lines. 
Comparing models M1-M4 in Figure 8, the overall trend observed with regard to the {\em total} accumulation of mass illustrated in Figure 7 also more or less holds for individual fragments: the mass accumulated by the primary fragments grows while the mass contrast with the secondary fragments tends to increase when the perturbation level goes up. This conclusion is clearly evident for model M4.  

Figure 8 also investigates the trend seen in Figure 7 for models M2, M5 and M6 at the level of individual fragments. Although the total mass of the primary fragments is larger at higher temperatures, it appears that this is not necessarily true for individual fragments in contrast to models M1-M4. In models M5 and M6, for example, a secondary fragment ends up having the biggest mass at the end of the simulation, although the total mass of the primary fragments starts to dominate in these models and is already equal to the mass contained with the secondary fragments in model M5. This trend can be explained by the fact that higher temperatures make it harder for the core disk to fragment into a high number of clumps and also delay the fragmentation process. When the fragmentation of the disk finally persists, the few emerging secondary fragments still have a considerable reservoir of material left to accrete from the disk, despite the initial advantage of the fragments resulting from the initial perturbation of the cloud. That is why some secondary fragments are still able to overtake the primaries in terms of mass accretion.

An important difference between our work and the recent simulations by Lomax et al. (2014) is the formation of brown dwarfs in their models, which is especially pronounced in the absence of radiative feedback. We attribute these differences to the different initial conditions because \citet{b300} have employed a turbulent initial gas cloud whereas we considered rigidly rotating cloud models without turbulence. We also emphasize that the initial conditions for star formation are not sufficiently
understood and need to be explored further in future work.    
  
Numerical simulations of star cluster formation reported by \citet{b28} confirm that protostellar disks formed around protostars are gravitationally unstable and prone to the development of spiral density waves. In our models of collapsing isolated solar mass cores, we also observe protostars with associated spiral density waves. In all examined models, however, these spiral density waves are found to be associated with close binaries which are still accreting mass from their surroundings. However, isolated protostars with gravitationally unstable disks do exhibit a clumpy spiral structure but only at an earlier stage of their evolution. As these isolated protostars evolve and accrete matter from their surrounding disks, these spiral structures disappear when the mass of their disks declines and the gravitational instabilities are suppressed. Within our simulations, the presence of binary systems thus tends to correlate with the presence of a self-gravitating disk. Such a behavior can potentially make sense, as a larger gas reservoir may be necessary for the formation of a binary system, which may subsequently favor the formation of a self-gravitating disk. We note, however, that statistically relevant conclusions cannot yet be drawn at this point.

\subsection{Bursts and accretion}

Figures 9 and 10 illustrate the history of the accretion rate for the individual protostellar fragments in simulations M1-M3 (Figure 9) and M4-M6 (Figure 10),    respectively.
Both Figures 9 and 10 show that the accretion rates initially peak due to the accretion of low-angular momentum material right at the stage of protostar formation, and then on average decrease over time, with typical average accretion rates of about 10$^{-6} M_{\odot}/yr$ within a timescale of 10$^{5}$ years. Even lower accretion rates seem possible at later evolutionary stages.  Moreover, some sink particles are gravitationally ejected from the gas cloud so that accretion onto these protostars is suppressed. In addition to these global trends, individual peaks as well as short-term variations are superimposed on the long term trend. These peaks are the hallmark of episodic accretion.

The first spike in the mass accretion profiles indicates the event of protostar formation. The first spike is then followed by a second spike which indicates the first and sometimes only accretion burst. Protostars evolving as individual objects in our simulations show accretion burst(s) at earlier stages of the cloud collapse. For instance, in model M1, sink 6 (secondary 4), in model M2, sink 1 (primary 1) and sink 4 (secondary 2), and in model M3, sink 2 (primary 2), have undergone their first or only accretion burst after 45 kyr of molecular core evolution. The curves which show the accretion history of isolated protostars in Figure 9 (a green line in top-left panel and a cyan line in top-right panel; a blue line in middle-left panel, and blue, green \& red lines in middle-right panel; a green line in bottom-left panel) and in Figure 10 (green lines in top-left and top-right panels; a blue line in bottom-right panel) show that for the solar mass molecular cores simulated in this paper, protostars formed in isolation and represented by isolated sink particles only produce accretion bursts when these protostars are initially part of a compact system with small distances between the protostars ($\le$ 500 au) and a considerable gas reservoir available for accretion. While being part of a strongly interacting system at the initial stages of their evolution, these objects have enough material in their immediate surroundings to produce episodic accretion bursts. An illustration of this phenomenon can, for instance, be found in model M1 where the sink 2 (primary 2) never shows any sign of episodic accretion, while the isolated protostars sink 1 (primary 1) and sink 2 (primary 2) in models M2 and M3 do show strong episodic accretion but only at the earlier stages of the core collapse (i.e. when the system is more compact and strongly gravitationally interacting). For example, the accretion bursts observed in sink 1 (primary 1) in model M2 cease after about 53 kyr. Sink 2 in  model M3, on the other hand, continues to show accretion bursts until 90 kyr but the strength of these episodic accretion bursts keeps decreasing over time. As the collapsing cloud evolves further, these sinks become quiet and show either no or weak spikes in their mass accretion history. Furthermore, the accretion history of the sink particles included in binary systems in models M1-M4 (as shown in table 4) suggests that episodic accretion bursts can also happen in binary systems as revealed by the spikes in their rate of mass accretion. The most important difference with isolated protostars is that for binaries episodic accretion can continue until much later stages of core collapse as illustrated by the curves for binary components in Figures 9 and Figure 10. In addition to the short-term accretion events that temporarily enhance the accretion rate, the longer-term trends show a decline of the average accretion rate with time. There is clearly a connection with the decreasing mass in the envelope. Only in the presence of a sufficiently big mass reservoir, the accretion onto the protostars can be maintained. The depletion of the envelope could explain why the observed accretion luminosities of some protostars are lower than expected, if we assume that most of the accretion has already happened during the embedded phase of the protostars which lasts for about half a million years, whereas the entire protostellar phase takes around a million years. 
   
Another important observation from our simulations is that there does not appear to be any kind of synchronization between the accretion bursts occurring in the two components of binary systems at the earlier stages of cloud collapse. However, once these objects continue to evolve as isolated binaries within the collapsing gas, they indeed start showing precisely synchronized bursts but with a relatively weaker amplitude. 
This feature can be explained if these binaries end up evolving as individual systems (hence without being fed by gas which is being channeled through gravitational interactions with a nearby protostar(s)) moving through the ambient gas from one point to another. The surrounding gas acts as a uniform source of mass accretion for both components constituting the binary system which has evolved to high mass ratio (q $\ge$ 0.75) by this time such as sink 1 (primary 1) \& sink 3 (secondary 1) in model M1, sink 2 (primary 2) \& sink 4 (secondary 2) in model M2, sink 1 (primary 1) \& sink 3 (secondary 1) in model M4 as can be noticed in table 4. One of the binary systems formed in model M3, however, is an apparent exception to this rule. The binary composed of sink 1 (primary 1) and sink 5 (secondary 3) in this model exhibits synchronized episodic accretion bursts despite its much lower mass ratio of 0.499.  

We suggest that the time of occurrence of episodic accretion for isolated protostars is controlled by how long the system hosting the individual protostar remains a dynamically compact structure. We also suggest that this mainly temporal constraint for episodic accretion is relaxed for VLM binaries. In these binary systems the non-steady accretion flow is controlled by spiral waves around the binary systems which transfer angular momentum outward and gas inward as explained by \citet{b30}. No such supply of material is possible for individual VLM stars evolving in isolation.

In models M2, M5, and M6, where the initial temperature of the cloud cores is increased in steps of 2 K, we do observe stronger and more frequent spikes in the mass accretion rate at the later stages of evolution even for binary systems including two secondary components. Taking into account the Bondi-Hoyle accretion rate, \citep{b301}

\begin{equation} \label{accretion rate}
\dot M_{BH}=\frac{4 \pi G^{2}M^{2}\rho_{\infty}}{(c_{\infty}^{2}+v_{\infty}^{2})^{3/2}},  
\end{equation}

which is more relevant to the case of the disk with multiple fragments, indicates that the accretion on the (secondary-secondary) protobinary system in model M5 is not governed merely by the prevailing sound speed which is relatively high because of the hotter core temperature compared to what we have in model M2. Despite the expected low accretion rate the (secondary-secondary) protobinary system in model M5 exhibits accretion spikes even at the later stages of the collapse. A potential explanation could be the location of this protobinary system inside the gas envelope. The local density of the gas plays a role in determining the accretion rate for the protostars \citep{b302}. Also, Bondi-Hoyle-Lyttleton accretion becomes a more significant mechanism when the mass of the stars grows.

Our simulation results lead to the interesting conclusion that the presence of a matter-rich surrounding around individual protostars can give birth to spiral density waves which lead to accretion bursts, as described in \citet{b3}. There is no obvious dependence on binary parameters such as the binary mass ratio, the eccentricity, the total mass contained within the binary system, as well as core parameters such as the ratio of the envelope mass to the initial mass of cloud core as discussed in table 3 and 4. 

Our simulations show an interesting correlation between the occurrence of episodic accretion and the mass depletion of the envelope. In the initial phase of model M1, before the envelope has lost 40 percent of its initial value in around 50 kyr, we observe much more frequent and stronger accretion bursts. Their amplitude and frequency drops down considerably as we move forward in time. So when the accretion from the envelope drops down, the amplitude and frequency of the accretion bursts also falls down significantly. However, as we increase the initial density perturbation of the cores in models M2, M3, and M4, we observe similar strong and frequent accretion bursts before the stage of 40 percent mass depletion of the envelopes in these cores, but the bursts continue to occur even after this stage has been reached. This indicates that the protostars formed in higher initial density perturbation cores have disks around them which remain dominant enough to support episodic mass accretion for protostars that are components of binary systems. These preliminary trends should be tested in a set of simulations exploring a larger range of initial conditions.

       \begin{figure*} \label{fig:11}
    \centering
    \includegraphics[angle=0,scale=0.550]{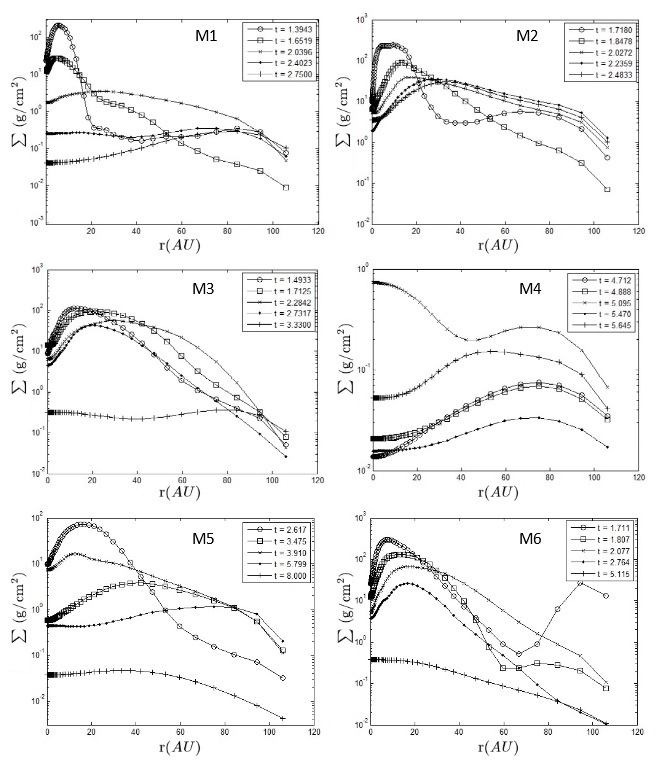}
  \caption{Radial surface density profiles associated with sink 1 for models M1$\rightarrow{}$M6. The extent of the radius in physical units is 120 au.}
     \end{figure*}

\begin{figure*} \label{fig:12}
    \centering
    \includegraphics[angle=0,scale=0.550]{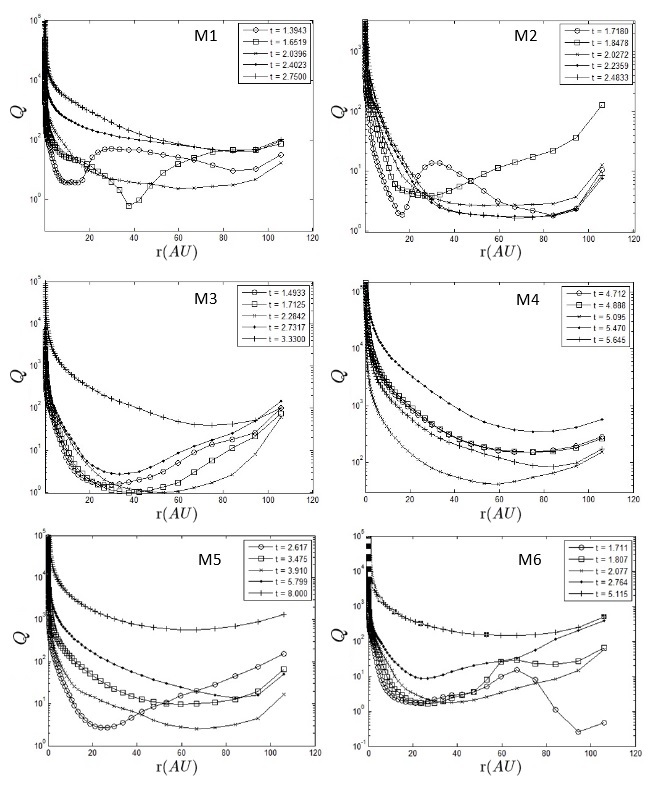}
    \caption{Traces of the azimuthal average of the Toomre parameter values as a function of the radial distance in the disks associated with sink 1 for models M1$\rightarrow{}$M6. The extent of the radius in physical units is 120 au.}
     \end{figure*} 
     
     
    \begin{figure*} \label{fig:13}
    \centering
    \includegraphics[angle=0,scale=0.55]{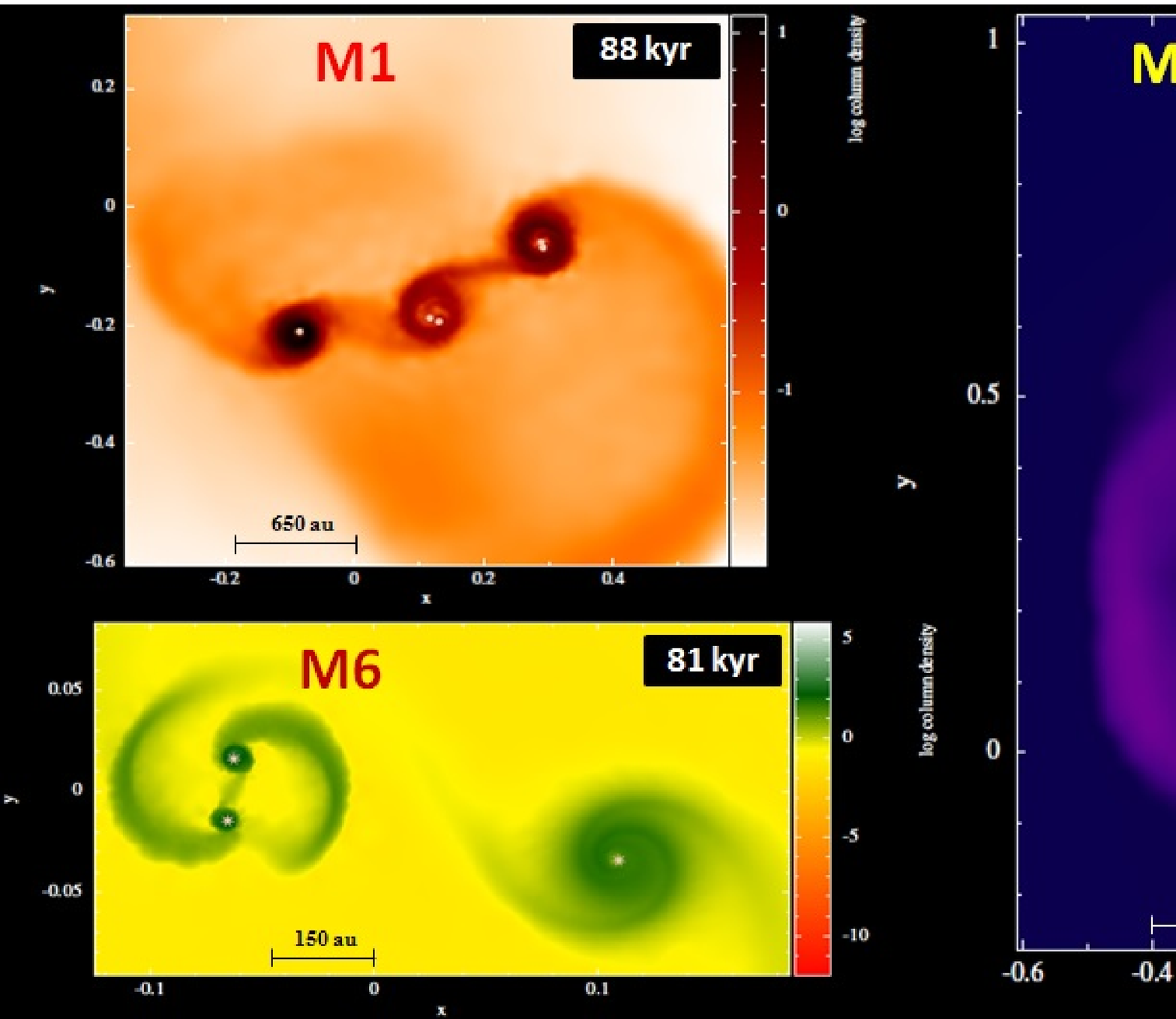}
    \caption{Column density plots which illustrate typical spiral structures observed during the evolution of models M1, M4 and M6, respectively. The physical sizes of each respective panel are 334229 x 317518 au, 267383 x 434498 au, 108625 x 66846 au.}
     \end{figure*} 

\subsection{Disk accretion} 
 
We now investigate the properties of the circumstellar disks orbiting the sink particles in more detail. In order to identify the SPH particles which are part of the disks, we require that those particles meet the first 3 criteria defined in \citet{b49} which require that disk particles have a nearly Keplerian velocity with respect to the sinks, and are part of a disk which is nearly in hydrostatic equilibrium and rotationally supported. We do not impose a threshold on the density of particles included in disks.

Figure 11 shows the azimuthally averaged radial profiles of the surface density of the gas distribution near the primary fragment sink 1 for the set of models (M1$\rightarrow{}$M6). The temporal variation of the surface density profiles indicates that the disks continuously accrete matter from their surroundings. In all models (M1, M2, M3, M4, M5, M6), sink 1 is a member of a protobinary system. Choosing sink 1 to analyze the disk accretion allows us to study the behaviour of protobinary disks for all initial conditions examined in this paper. We try to compare the effect of changing the initial temperature of the cores and changing amplitudes of initial azimuthal density perturbation on the evolution of the disks.

Figure 12 shows the azimuthally averaged radial Toomre Q parameter profiles for the disks related to the primary fragment sink 1. In model M1 the Q parameter profiles suggest that the disk remains stable for most of the time but at an earlier stage of the evolution around 1.487 $t_{ff}$ the Q parameter drops below unity in the radial range around 40 AU. The drop in the Q parameter indicates the possible start of fragmentation within the disk. On the other hand, gravitational instabilities within the disk orbiting sink 1 in model M2 are far weaker. However, model M3 shows a quite different behaviour. In model M3 the disk associated with sink 1 continues to develop gravitational instabilities for a wide range of disk radii (from 20 au to 60 au). This behaviour lasts between 1.344 $t_{ff}$ to 2.056 $t_{ff}$ but later on the disk becomes stable as can be seen in the middle-left panel of Figure 12. Models M5 and M6 are stable against gravitational instabilities throughout their entire evolution. Model M6, however, shows an interesting feature related to the possible occurrence of gravitational instability at a much larger radius within the disk. Unlike in any other model, the disk around sink 1 in model M6 develops a gravitational instability at a much larger radius around 100 AU. Later on the disk associated with the sink loses mass and regains gravitational stability. 
 
Figure 13 represents column density plots which illustrate the presence of spiral structures which form in the vicinity of fragments that belong to a binary or a triple system or are in the process of accreting material from their parent core as individual clumps. As we discussed previously, the spiral trail of gas around single fragments is a feature which is observed only during the earlier evolution of the system. As the collapsing molecular core evolves further in time, the spiral structures seen around single fragments gradually disappear as the disks are losing mass by accretion onto the sink and are becoming stable to gravitational fragmentation. In addition to this, it is at present difficult to establish a connection between the outbursts seen in Figures 9 and 10 and the spiral structures seen in Figure 14 as gravitational instabilities within single disks cannot explain the outbursts, but tidal effects by passing fragments and waves generated by orbiting binaries are a better candidate. 

\subsection{Binary properties} 
  
Recent observational studies of very low-mass binaries have revealed a broad range of eccentricities (0 $\rightarrow{}$1) associated with these systems \citep{b44}. The orbits of protobinaries which have been formed in our simulations via relatively later disk fragmentation (of the secondary-secondary type) tend to be more circular in our models. However, if at least one of the companions of such a system is a primary fragment that is likely formed even before the core disk structure has clearly emerged, the resulting protobinary (of the primary-secondary type) appears to have a highly elliptical orbit. This trend continues towards higher eccentricities if both components of a protobinary result from initial fragmentation (the primary-primary type) as found in model M5. However, the protobinary of the primary-primary type in model M6 remains an exception in that it appears to have a much lower value of the eccentricity and tends towards a circular orbit.  
Observational studies have also revealed that low-mass binaries at small separations have reduced eccentricities \citep{b47}. This feature can possibly be explained as the result of tidal circularization if these binaries have been formed from fragments in a disk and continue to tidally interact with material in the disk \citep{b47}. Table 4 clearly shows that our models M1, M2, M3, and M5, which have produced low-mass binaries of the secondary-secondary type, indeed have a small separation and a reduced eccentricity. Both companions in each binary are of later type fragmentation and hence they could have experienced tidal circularization. For wide binary orbits tidal circularization may be a negligible effect\citep{b48}.

Unlike the solar type population of binaries, observations reveal that there exists no correlation between eccentricity and period or semi-major axis when samples of low-mass binaries are examined \citep{b44}. Our simulation findings are in good agreement with these empirical results as table 4 shows that there is no correlation of the period with the eccentricity and the semi-major axis in the simulation models that yielded low-mass binary systems. It is also interesting to note that the observed binary properties have been found to be inconsistent with previous hydrodynamic simulations \citep{b44,b45} with regard to the binary eccentricities. We suggest that the model of an isolated self-gravitating solar mass core can bridge the gap between observations and theory with regard to the eccentricity of low-mass binaries. Unlike a cluster environment, an isolated solar mass molecular core gravitationally collapses to form multiple low-mass protostellar systems and a whole range of protostellar binaries with varying eccentricities as reported in the present paper. More importantly, unlike the too many highly eccentric orbits produced by the simulations reported in \citep{b46}, our models have produced binaries with a wide range of eccentricities (0.056 $\le$ e $\le$ 0.960) which is consistent with the results of \citep{b44}. However, the possible impact of small number statistics means that these results are at best preliminary and more extensive simulation studies will be needed in future work.

  
\section{Conclusions and outlook}

The simulations reported in this paper demonstrate that VLM stars can be formed as a result of both primary fragmentation as well as secondary fragmentation within core disks formed from collapsing solar mass molecular cloud cores. The collapsing cores form a variety of protostellar binaries and triple systems. 
In contrast to weakly perturbed clouds, strongly perturbed clouds produce fewer fragments within a limited range of masses. In addition, the level of the density perturbations also seems to control the delay between primary and secondary fragmentation. 

We find that warmer cores produce core disks which become depleted and fall apart in circumbinary and circumstellar disks. The accumulation of mass associated with different types of fragments shows several trends. Going from small to large initial density perturbations, primary fragments are most likely to lead the mass accumulation process. However, when the cores become warmer, individual secondary fragments can eventually become more massive than the primary fragments although the total mass accumulated within these fragments remains dominant. We overall find that the accretion rates show a short peak at the time of formation, which then follows an average trend of decreasing accretion rates, due to the depletion of the gas from the envelope. This average development is superimposed with peaks and short-term variations in the accretion rate, as a result of episodic accretion.

Protostars which are formed as single objects within isolated solar mass cores can exhibit accretion bursts in our models only when they are part of a matter-rich environment of a multiple system at the earlier stages of collapse. However, the possibility of burst-like accretion indeed arises even later and in isolation if a solar mass core hosts protobinary systems, but seems to be suppressed again if the cores become warmer. Despite clear evidence of the association of accretion bursts with protobinaries, so far no connection has been found with the orbital elements of the binaries, hence preventing the prediction of the time of occurrence of the accretion bursts.

We find that protobinary systems consisting of (primary-primary) fragments are highly eccentric and with stronger episodic accretion. If one of the companions is of the secondary type then the eccentricity tends to get reduced. Interestingly, if protobinaries are of the secondary-secondary type, hence resulting from later disk fragmentation, the eccentricity is reduced further and the systems becomes more circular. This can likely be explained by the presence of tidal interactions with their surrounding circumbinary disks which can lead to circularization of the system. However, if the temperature of the collapsing parent core is increased further to 12 K, even the protobinary systems of (primary-primary) type can become less eccentric (\textit{e} $<$ 0.4). As emphasized in various places, the putative trends we found in the simulations performed here should be investigated employing a larger range of realizations and initial conditions. Exploring the difference between different types of initial conditions, including turbulent gas clouds, will also be important in the future. The latter will be valuable to assess the uncertainties and spread in the model predictions, whether preliminary trends we find here hold also in larger samples, and whether different types of initial conditions can lead to a bias in the results.

It will also be important to explore the impact of the equation of state on the collapse dynamics and the fragmentation process. As shown by \citet{b56, b57}, fragmentation occurs through the formation of self-gravitating disks if the gas temperature increases with density, while a decreasing temperature for increasing density leads to a filamentary fragmentation mode. The latter was shown to be relevant for instance at low metallicity, when dust grains contribute to the cooling in the primordial gas \citep{b58}. Extending the studies pursued here to different regimes may thus shed additional light on episodic accretion events, fragmentation and the formation of filaments under different conditions.

In addition, future studies may need to explore in further detail the formation of stars within filaments. The integral shaped filament in the Orion A cloud shows strong indications for a dynamical heating of the stars, which can possibly be explained by an oscillation of the filament \citep{b68}. The latter scenario is indeed consistent with stellar-dynamical calculations \citep{b69}. Such filamentary structures may substantially affect the time-dependence of the accretion rate, and the evolution of the protostars.

\begin{acknowledgements}

This research has made use of the high performance computing cluster at the Abdus Salam Centre for Physics, Pakistan. The first author (R.R.) gratefully acknowledges support from the Department of Astronomy, University of Concepcion, Chile. The first author RR and the third author DRGS thank for funding through the Concurso Proyectos Internacionales de Investigaci\'on, Convocatoria 2015" (project code PII20150171). DRGS further thanks for funding via Fondecyt regular (project code 1161247) and via the Chilean BASAL Centro de Excelencia en Astrof\'isica yTecnolog\'ias Afines (CATA) grant PFB-06/2007, and ALMA-Conicyt project (proyecto code 31160001). The second author (S.V.) developed the computer code which has been used in this work on the HPC system Thinking at KU Leuven, which is part of the infrastructure of the FSC (Flemish Supercomputing Center). We thank the KU leuven support team for helping with the use of this system. The second author also gratefully acknowledges the support of Prof. Dr. Stefaan Poedts and Prof. Dr. Rony Keppens for having provided both access and funding which made the use of the KU Leuven HPC infrastructure possible.
\end{acknowledgements}

%
%

\end{document}